# Extraordinary optical transmission and vortex excitation
# by periodic arrays of Fresnel zone plates

A. Roszkiewicz and W. Nasalski*

Institute of Fundamental Technological Research, Polish Academy of Sciences
Adolfa Pawińskiego 5b, 02-106 Warsaw, Poland
arosz@ippt.gov.pl, wnasal@ippt.gov.pl

**Abstract.** Extraordinary optical transmission and good focusing properties of a two-dimensional scattering structure is presented. The structure is made of Fresnel zone plates periodically arranged along two orthogonal directions. Each plate consists of two ring-shaped waveguides supporting modes that match the symmetry of a circularly polarized incident plane wave. High field concentration at the focal plane is obtained with short transverse and long longitudinal foci diameters. Optical vortex excitation in a paraxial region of the transmitted field is also observed and analysed in terms of cross-polarisation coupling. The structure presented may appear useful in visualization, trapping and precise manipulations of nanoparticles.

**Keywords.** Fresnel zone plates, focusing, cross-polarisation coupling, optical vortices

## 1. Introduction

Discovery of extraordinary optical transmission (EOT) in a periodic set of holes drilled in opaque metal layer [1] stimulated extensive research on high transmission through plasmonic structures [2-15]. Different configurations were proposed like one-dimensional slits [2-4] or two-dimensional structures of rectangular [5] and circular symmetry [6-15], pertaining empty holes [6-7] and coaxial guides [8-11]. In addition, different incident field polarizations - linear [12], circular [13], azimuthal [14] or radial [15] - were taken into account. The goal of research was to obtain high transmission with possibly long focus of short transverse diameter.

On the other hand, manipulations of optical information carried in optical field amplitude, phase and polarization become also an important issue. Within this range the generation of an orbital angular momentum (OAM) was invented in several different ways, based mainly on direct changes of OAM introduced in the field helical phase fronts, on spin-orbit conversions by enforcing changes in the field spin angular momentum (SAM), or on simultaneous manipulations on both OAM and SAM of vector beam fields [16-21]. The spin-orbit conversion also occurs through the cross-polarization coupling (XPC) in fields of vector beams impinging on planar, transversally homogeneous media interfaces or layered structures [22]. In general, the XPC interactions always exist for beam reflection, refraction and trasmission under arbitrary incidence [23], even in cases without the spin-orbit conversion [24].

The goal aimed in this paper is placed within these two areas of interests. Interaction of a monochromatic, circularly polarized plane wave with a two-dimensional periodic array of two-ring Fresnel zone plates (FZPs) is numerically analysed and graphically described. The case of normal incidence is considered, in parallel to the recent report on EOT through a single-ring coaxial structure [15]. Field intensity, phase and polarisation distribution will be discussed in detail, with special attention given to the region near the field focus. The presence of optical vortices, as anticipated by analogy to the analysis given in [22], will be checked.

Two distinct phenomena are observed. First of all, as lateral dimensions of the structure (radii of the rings) are chosen accordingly to the FZP rule, efficient focusing of the transmitted field is obtained. The FZP alone focuses the field below the structure, even under the plain incident plane wave. Contrary to the case considered in [15] no additional prefocusing of the optical field is needed. The field focus is characterized by a narrow transverse cross-section and a long longitudinal length. The second effect of the structure appears to be the modification - in a well ordered manner - the field polarization state, as well as its phase and amplitude distribution. This feature is especially enhanced in the paraxial region close to the field focus. In effect, vortex excitation in the transmitted field exists and is clearly visible. Main charactestics of the vortex follow those known from the beam XPC interactions with planar layered structures [22].

The paper is organized as follows. Section 2 contains description of geometry of the FZP structure. The focusing properties of the structure are presented in Section 3. The amplitude and polarization distributions in horizontal planes inside and outside the structure are shown in Section 4. The excitation of modes in coaxial waveguides of the FZP, the XPC interactions and the presence of optical vortices in the orthogonal – to the incident - field components are discussed. In Section 5 field polarization and energy flow distributions are described near the field focus. Finally, Section 6 concludes main results of the paper.

## 2. Geometry of the structure

Geometry of the problem is presented in Figure 1. A right-circularly polarized (CR) plane wave of wavelength $\lambda = 876\ nm$ impinges on an infinite, periodic, two-dimensional structure under normal incidence. The structure, surrounded by air, consists of two-ring Fresnel zone plates etched in a lossless perfect electric conductor (PEC) layer with a dielectric constant $\varepsilon_{PEC} = -10^8$. The case of PEC was chosen in this analysis to minimize light interaction between air rings in one FZP and between neighbouring PZPs. No $E_z$ field exists inside the perfect conductor due to boundary conditions requiring vanishing of the tangent E field component at the PEC boundary. Electric field is confined inside the etched air rings. Periodicity of the structure has no influence on the field inside the neighbouring air rings and should only cause interference in the transmitted field below the structure. Moreover, the comparison of the FZPs etched in real metal (silver) is also presented in Figure 6 below and main differences in the field distribution between the PEC and silver structures are indicated.

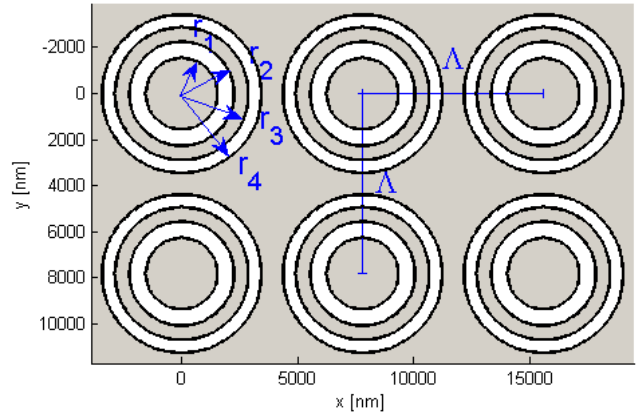

**Fig. 1.** Geometry of the problem: top view of the two-dimensional periodic structure consisting of two-ring Fresnel zone plates.

Accordingly to the Huygens principle, all points in the air rings are considered as sources of secondary spherical waves. Then, the constructive interference in the focal point yields the distance between the edge of each zone and the focal point $R_m = f + m\lambda/2$, where f is the focal length and $m=1,2, \dots$ This leads to the equation for inner and outer radii of each ring in FZP: $r_m = \sqrt{\lambda f m + (\lambda m/2)^2}$, where $m = 1,\dots,4$ [25]. The focal length is assumed to be $f = 2500\ nm$, which gives: $r_{1-4} = 1543, 2269, 2880$ and $3439\ nm$. Thickness of the structure ($d = 600\ nm$) and the limited number of the rings in each FZP slightly move the focus from the theoretical value to $z = 2980\ nm$. Periods in x and y directions are equal $\Lambda = 7800\ nm$ and allow for propagation of $+/-8$ diffraction orders in air at a given wavelength.

## 3. Focusing by the structure

Numerical analysis of the optical response of the structure is accomplish by using the rigorous coupled wave analysis [26] adopted to 2D structures [27] with implementation of the scattering matrix algorithm [28] and the factorization rules [29-31]. In the following calculations 21 diffraction orders in each transverse directions were assumed, which allowed for numeric operations on matrices consisting at most of $(4\cdot21\cdot21)^2 = 3111696$ elements.

Figure 2 presents the intensity of the total transmitted electric field in the vertical (a) and transverse (b) cross-sections of the focus. The figure confirms that the focusing ability of the structure is high in spite of that it consists only of two air rings. The diameter of the total electric field intensity in transverse direction at the focal plane, defined as the full width at half maximum (FWHM), is approximately equal to $0.57\lambda{\sim}500$ nm. It is lower by 36% than the Rayleigh resolution $R_{Rayl.}^{FZP} = 0.61\lambda/NA = 1.22\Delta r_4$ defined here for the FZP consisted of two air rings, which is equal 682 nm.

The structure allows also for the large length of the focus in the propagation direction. The focus length, where the electric field intensity does not fall below 50% of its maximum intensity, expressed in the incident wave number k is equal $l_{foc} \approx 2\pi w_w^2/\lambda \approx 448.29k$, which gives 3400 nm. The length of the focus for FZP can be estimated by $l_{foc} = \pm \lambda/[2(NA)^2] = \pm 2(\Delta r_4)^2/\lambda$ [32]. Here, $l_{foc}$ is defined as a distance from the focal plane, at which the intensity on axis is diminished by 20%. Accordingly to this equation $l_{foc} = \pm 713nm$, which is slightly larger than numerically calculated (±630nm).

Thus, the configuration allows for focusing plane wave to subwavelength dimensions with long focal depth at a certain distance from the structure. This configuration does not need further prefocusing of the optical field, as it was applied in [15]. In that case the optical beam was focused before illumination of a single coaxial aperture. The total transmittance, in reference to the transmittance through a glass substrate (*n=1.5*), was measured. The measured normalized transmittance for radially and azimuthally polarized beam was around 0.45 and 0.08, respectively. Thus the highest measured transmittance corresponds there to highly focused Gaussian beam of radial polarization and is around 0.43. In our configuration, without the need of prefocus of Gaussian beam, the total transmittance through structure is T = 0.41 for PEC. Additional difference is the case of circularly polarized plane wave discussed here.

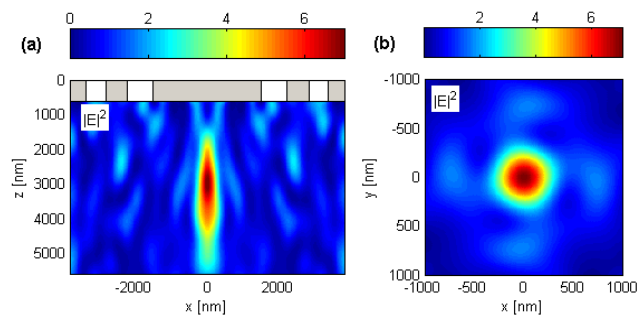

**Fig. 2.** (a) Side view (cross section at y = 0) of the total electric field intensity distribution below the grating with one period visible. (b) Total electric field intensity distribution at the focal plane (z = 2980 nm) in the paraxial region.

Note that EOT occurs when the normalized-to-area transmittance is larger than unity [33]. The case analysed here can be considered as EOT since the normalized-to-area transmittance (defined as the transmission normalized to the amount of light impinging on the area occupied by the air rings) is 1.25.

## 4. Field spatial distribution

Figure 3 presents total, transverse and longitudinal electric field intensity distributions in the middle of the structure (z = 300 nm), just below the structure (z = 601 nm) and at the focal plane (z = 2980 nm). The transmitted field consists of the superposition of waves diffracted from several neighbouring FZPs, but the main output in the vicinity of the z-axis of each FZP structure is originated only from the one FZP placed just above. Moreover, the calculation of amplitudes of the subsequent diffraction orders indicate that the amplitude magnitude of the zero (0,0) transmitted order is at least three times larger than the one of any higher order. Therefore, we are mainly interested in the polarization and phase changes around the axis of each structure, that is in the paraxial range near the field focus. Note that the relation between the wavelength and horizontal dimensions of the structure is $2\pi r_i/\lambda > 10$ for each i=1,...,4. Because all of that we present below the field and phase distributions only near the FZP axis in the following Figures (4-10).

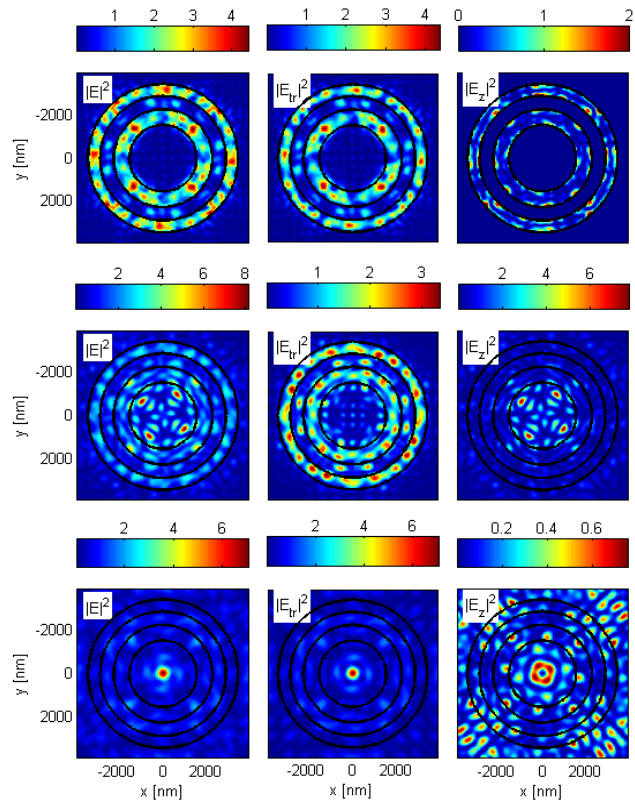

**Fig. 3.** Electric field intensities; first column: total field, second column: transverse field, third column: longitudinal field. First row: in the middle of the structure at z = 300 nm, second row: below the structure at z = 601 nm, third row: in the focus at z = 2980 nm. The concentric rings of the structure are indicated. Numerical calculations with only 21x21 diffraction orders are taken into account. Thus the positions of the maxima around the air rings are approximate, not exact.

In Figure 3 strong focusing ability of the structure is revealed for all, the transverse and the longitudinal, field components. The characteristic $E_z$ field distribution with maximum field intensity around the z-axis and zero at the centre is clearly visible not only in the focal plane, but even just below the structure, at z = 601 nm.

Regularly placed maxima are visible in transverse electric field components inside the coaxial rings. They are originated from the superposition of the excited waveguide modes in the structure. Since the normally incident plane wave has circular polarization, the symmetry does not allow for basic TEM$_{00}$ mode excitation [34], the only mode without cut off wavelength in this configuration. Other waveguide modes, to be excited, have to match the symmetry of incident circular polarization and fulfil the boundary conditions. An amplitude and phase of electric field vector of any mode in real metal needs to fulfil the Bloch theorem. However, in the case of perfect metal there is no coupling between the cavities and the eigenfrequencies of the modes are independent on Bloch vector [12]. The dispersion relation that determines the propagation constants of modes in real metal are presented in Appendix.

The modes excited inside the concentric waveguides are rather hybrid modes than pure TE or TM modes, since both electric and magnetic fields have nonzero longitudinal components $E_z \neq 0$ and $H_z \neq 0$ at each cross-section plane inside the guide. The analytical expressions for electric field components for TE and TM modes and the cut off wavelengths $\lambda^{co}$ for TE and TM modes in a coaxial waveguide are given in [35]. In the case of four first modes in the inner and outer rings $\lambda^{co}$ is approximately equal to: 11976 nm and 19854 nm for TE$_{11}$ mode, 5988 nm and 9927 nm for TE$_{21}$ mode, 2279 nm and 1756 nm for TM$_{11}$ and TM$_{21}$ modes, respectively. The number of maxima around the outer air ring indicates that also modes of higher indices are excited. These are examples of modes that matches the polarization symmetry of the incident wave and can be excited in this case.

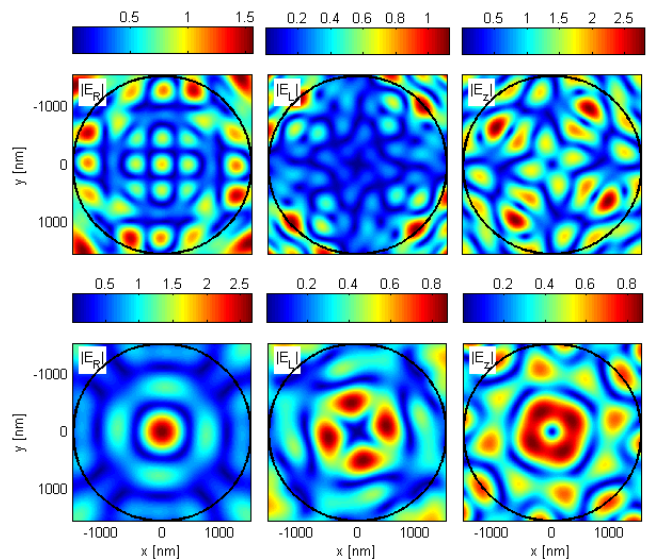

**Fig. 4.** Absolute amplitude distributions of transverse $|E_R|$, $|E_L|$ and longitudinal $|E_z|$ electric field components for structure made of PEC. First row: below the structure (z = 601 nm). Second row: at the focal plane (z = 2980 nm). Note that only the field in the paraxial region is presented.

More information about the interaction of circularly polarized plane wave and periodic structure made of FZP waveguides can be inferred from the amplitude and phase distributions (Figures 4 and 5). The amplitude distribution of direct (right circular (CR)) polarization has no sign of optical vortex excitation. The amplitude maxima are placed at the axis and in concentric outer ring, which is additionally modified by the grating periodicity $\Lambda$. The influence of neighbouring FZPs can be seen in the area beyond the focus spot. The field pattern indicates the diffraction of the propagating modes at the output of concentric apertures. Cross-polarized (left circular (CL)) and longitudinal components, on the other hand, exhibit null amplitude at the axis and an outer ring of maximum amplitude, with four maxima around.

The phase distribution at the focal plane clearly indicates excitation of optical vortices with typical characteristics of the XPC interaction of vector beams of finite cross-sections with a planar layered structure [22]. The vortex of topological charge equal two is excited (twice $2\pi$-change of phase around the z-axis) in the transverse orthogonally polarized (CL) component and, consequently, also in the longitudinal field component with the toplogical charge equal one. In general, in such vortex excitations, the change of the optical toplogical charge by plus two (minus two) is accompanied by the switch of the field polarisation from CR to CL (CL to CR) in the horizontal plane z=const. [22].

Additionally, the phase of $E_z$ in an outer area around the focus axis is shifted by ~$\pi$ with respect to the phase of the focus region. Beyond that second phase ring this interference of waves incoming from neighbouring structures influences significantly the field and phase distribution. Similarly, a sharp phase changes at the edges of the focus spot are visible in the direct (CR) and orthogonally polarized (CL) transverse field components.

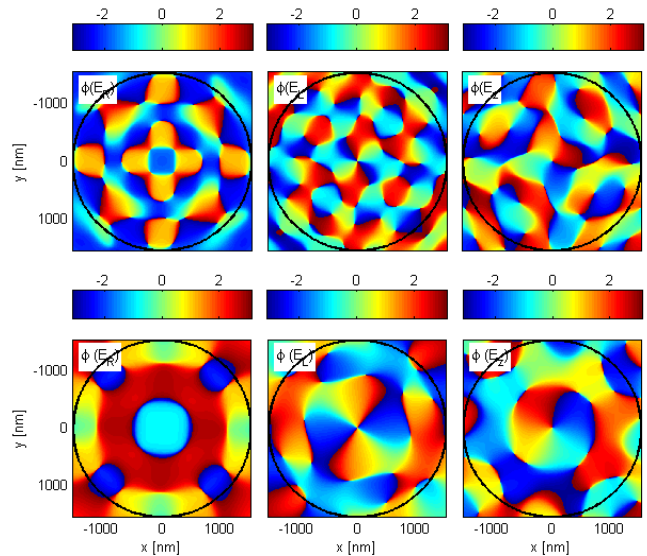

**Fig. 5.** Phase distributions of transverse $|E_R|$, $|E_L|$ and longitudinal $|E_z|$ electric field components for structure made of PEC, corresponding to amplitude distributions in Figure 4. First row: below the structure (z = 601 nm).

Second row: at the focal plane (z = 2980 nm).

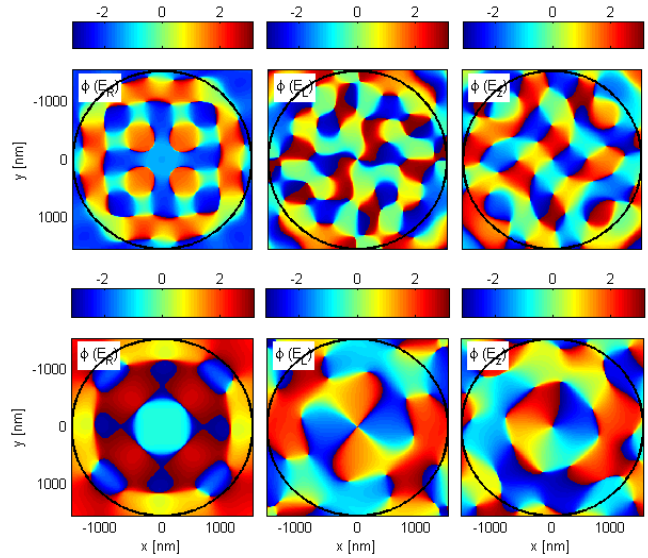

**Fig. 6.** Phase distributions of transverse $|E_R|$, $|E_L|$ and longitudinal $|E_z|$ electric field components for structure made of silver. First row: just below the structure (z = 601 nm). Second row: at the focal plane (z = 2980 nm).

It was shown in Figure 6 that the field distribution in real metal (silver with the dielectric function numerically fitted to experimental data [36]) is of the same sort as that described for PEC. Due to possible coupling between the periodically arranged guides in real metal, the presence of neighbouring structures modifies slightly the phase distributions with respect to PEC, particularly visible in direct polarization. Each FZP has four closest-neighbouring FZPs along x and y directions. The four disturbances visible at those directions around FZP indicate that week interaction between the closest FZP neighbours really exists.

The configuration of periodic air holes in the metal layer was also checked. In this case the vortex excitation - although week - is also possible. The field focusing, however, is much more pronounced in the case of the FZP structure considered here.

## 5.   Field polarization near focus

As it was numerically verified in Figures 4-6, the incident plane wave of CR polarization excites the phase singularity in the orthogonal CL polarization and, consequently, in the longitudinal field component as well. However, the field intensity of the CL field component in the region close to the focus remains in magnitude smaller than that of the CR polarization. Still, is interesting to see how this excitation influences the polarization distribution of the field near its focus.

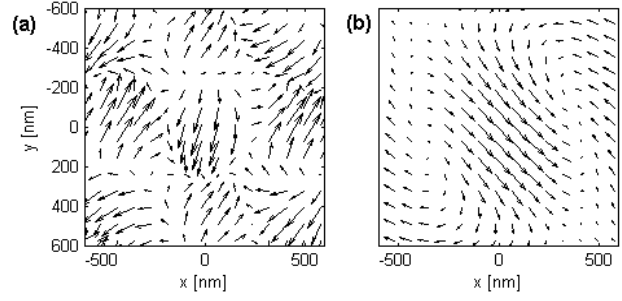

**Figure 8** Transverse real electric field direction (a) just below the structure (z = 601 nm) and (b) at the focal plane (z = 2980 nm). Note that only the paraxial region is presented.

Figure 8 presents the total transverse electric field direction denoted by arrows, just below the structure and in the focal plane. The parallelism of the arrows in the focus corresponds to the dominant right-handed elliptical polarization in paraxial region. The focusing of the field enlarges the range of quasi-homogeneous field polarization. However, the field polarization changes in the transverse plane of the focus. When the polarization vector $\underline{A}$ is decomposed into the x and y polarization components: $\underline{A} = \hat{x} A_x e^{i\delta_x} + \hat{y} A_y e^{i\delta_y}$ with the relative phase $\delta = \delta_y - \delta_x$, then the field in the rotated coordinates x' and y' is described by the ellipse: $\left(A_{x'}/a\right)^2 + \left(A_{y'}/b\right)^2 = 1$ with the ellipticity e=b/a and the inclination angle $\phi = 2^{-1} \tan^{-1}\left[2 A_x A_y \cos(\delta) / \left(A_x^2 - A_y^2\right)\right]$ [37].

Figure 9 describes the field polarization distribution near focus. The electric field polarization changes its handedness several times with distance from the focus centre, as it is clearly seen in Figure 9(a). Figure 9 (b) shows the polarization ellipse in the subsequent field points. In the centre point the circular polarization is exactly specified; the outer circle in Figure 9(b) corresponds to pure CR polarization at the axis with δ = π/2. As the distance from the focus centre grows, the polarization becomes more elliptical (dark-blue and light-blue ellipses) and at the point where sinδ = 0 the field is linearly polarized. The ring of linearly polarized field around the axis separates the areas where electric field polarization is right- and left-handed. Red ellipse corresponds to sinδ = -1, which means pure left-handed elliptical field.

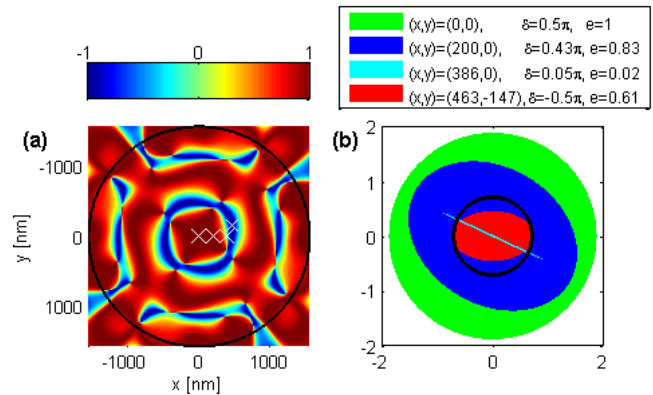

**Figure 9** (a) Values of sinδ in the paraxial region of the focal plane, where $\delta$ is the phase difference between two electric field components. (b) Polarization ellipses for four points denoted by crosses in (a). In the legend, subsequent values of δ and the ellipticity e are indicated. The black circle indicates the incident CR wave polarization.

The changes of the field polarization result in the circular flow of the optical field energy in the first intensity ring around the field focus centre. It is shown in Figure 10 within the paraxial range of the transmitted field near the focus. Separate regions of maximal values of transverse and longitudinal Poyting vector components confirm the optical vortex excitation in the centre of the beam field.

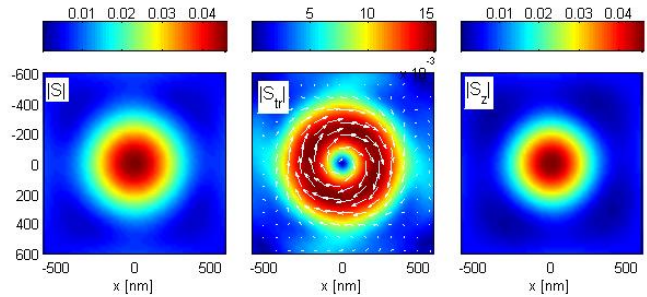

**Fig. 10.** **A**mplitudes (colour) and direction (arrows) of the Poyting vector $\underline{S}$ in the focal plane (z = 2980 nm); from left to right: total, transverse component, longitudinal component of the field in the paraxial region close to the focus centre.

## 6. Conclusion

A new type of focusing metallic structure is numerically analysed. It consists of periodically arranged FZPs engraved in perfect electric conductor or silver. The case of normal incidence of a circularly polarization plane wave is analysed. The structure allows for extraordinary optical transmission and efficient focusing with no need of prefocusing of the incident field. Moreover, the changes the transmitted field intensity, phase and polarization spatial distribution enforced by the FZP structure indicate the XPC interactions between polarisation field components [22]. That results in the optical vortex excitation in the orthogonal - to the incident one - polarization field components. The vortex diameter is narrow due to field focusing but still it does not prevail in intensity over the field of incidence polarisation. It seems, however, that the vortex excitation can be made stronger by the methods typical for resonant metamaterial structures [24]. Due to its properties the FZP structure may appear useful in visualization, trapping and precise manipulations of nanoparticles.

## Appendix

In the case of real metal (here silver) the theoretical calculations give the value of propagation constant equal to 1.0373 in the inner air ring and 1.0484 in the outer air ring for the first TM modes. The calculation rely on the assumption that the propagating mode in each air ring of the coaxial structure has the total wave vector determined by the wave vector for metal-insulator-metal (MIM) plane cavity [8]: $\beta^2 + k_\theta^2 = \beta_{MIM}^2$. Together with the condition $2\pi r_{ring} k_\theta = 2\pi\nu$ it gives the dispersion relation for an air ring in a coaxial structure:

$$\tanh\left(\sqrt{\beta^2 + \left(\frac{\nu}{r_{ring}}\right)^2 - k_0^2 \varepsilon_d}\,\frac{d}{2}\right) + \varepsilon_d \sqrt{\beta^2 + \left(\frac{\nu}{r_{ring}}\right)^2 - k_0^2 \varepsilon_m} \left/ \varepsilon_m \sqrt{\beta^2 + \left(\frac{\nu}{r_{ring}}\right)^2 - k_0^2 \varepsilon_d} = 0 \right. \qquad (A.1)$$

where $r_{ring} = (r_1 + r_2)/2$ for the inner ring and $r_{ring} = (r_3 + r_4)/2$ for the outer ring, $\nu$ is an integer, $\varepsilon_m$ and $\varepsilon_d$ - permittivities of metal and dielectric.